\documentclass[final,5p,times,twocolumn, authoryear]{elsarticle}
\pdfoutput=1

\usepackage[utf8]{inputenc}
\usepackage{lipsum}
\usepackage{graphicx}
\usepackage{natbib}
\usepackage[colorlinks=true,linkcolor=blue,citecolor=blue]{hyperref}%
\usepackage{hyperref}
\usepackage{amssymb}
\usepackage{amsmath}
\usepackage{nomencl}
\usepackage{siunitx}
\usepackage{cellspace}
\usepackage{array, makecell}
\usepackage{multibib}
\usepackage{units}
\makenomenclature

\usepackage{setspace}

\journal{Advances in Water Resources}

\begin{document}

\begin{frontmatter}

\title{The Impact of Sub-Resolution Porosity on Numerical Simulations of Multiphase Flow}
\author[PRINCETONCIVIL,*]{Francisco J. Carrillo}
\author[CNRS]{Cyprien Soulaine}
\author[PRINCETONCIVIL,HMEI]{Ian C. Bourg}

\address[PRINCETONCIVIL]{Department of Civil and Environmental Engineering, Princeton University, Princeton, NJ, USA}
\address[CNRS]{Earth Sciences Institute of Orl\'eans, Universit\'e d'Orl\'eans, CNRS, BRGM, Orl\'eans, France}
\address[HMEI]{High Meadows Environmental Institute, Princeton University, Princeton, NJ, USA}
\address[*]{Corresponding author: franjcf@outlook.com}
\date{\today}

\begin{abstract}

Sub-resolution porosity (SRP) is an ubiquitous, yet often ignored, feature in Digital Rock Physics. It embodies the trade-off between image resolution and field-of-view, and it is a direct result of choosing an imaging resolution that is larger than the smallest pores in a heterogeneous rock sample. In this study, we investigate the impacts of SRP on multiphase flow in porous rocks. To do so, we use our newly developed Multiphase Micro-Continuum model to perform first-of-a-kind direct numerical simulations of two-phase flow in porous samples containing SRP. We show that SRP properties (porosity, permeability, wettability) can impact predicted absolute permeabilities, fluid breakthrough times, residual saturations, and relative permeabilities by factors of up to 2, 1.5, 3, and 20, respectively. In particular, our results reveal that SRP can function as a persistent connector preventing the formation of isolated wetting fluid domains during drainage, thus dramatically increasing relative permeabilities to both fluids at low saturations. Overall, our study confirms previous evidence that the influence of SRP cannot be disregarded without incurring significant errors in numerical predictions or experimental analyses of multiphase flow in heterogeneous porous media. 

\end{abstract}


\end{frontmatter}

\section{Introduction} \label{sec:intro} 

The recent emergence of Digital Rock Physics (DRP) has revolutionized the way we study porous media. It is now possible to directly characterize the pore structure of subsurface systems and perform three-dimensional direct numerical simulations of fluid flow in digital models of rock samples that approach the size of a Representative Elementary Volume (REV). As such, DRP has transformed our capacity to characterize and predict fluid flow in soils, sedimentary rocks, hydrocarbon reservoirs, and engineered porous systems \citep{Mehmani2019a,Han2020}. The computation of rock transport parameters including absolute permeability \citep{Spanne1994}, dispersion coefficients \citep{Bijeljic2013a,Soulaine2021c}, relative permeabilities, and capillary pressures \citep{Raeini2014,Prodanovic2015} has had direct impacts in the fields of reservoir engineering, hydrology, and $\mathrm{CO_2}$ sequestration \citep{Blunt2013,Soulaine2021a}.

\subsection{Rock Imaging Techniques and Sub-Resolution Porosity}

\begin{figure*}[t!]
\begin{center}
\includegraphics[width=1\textwidth]{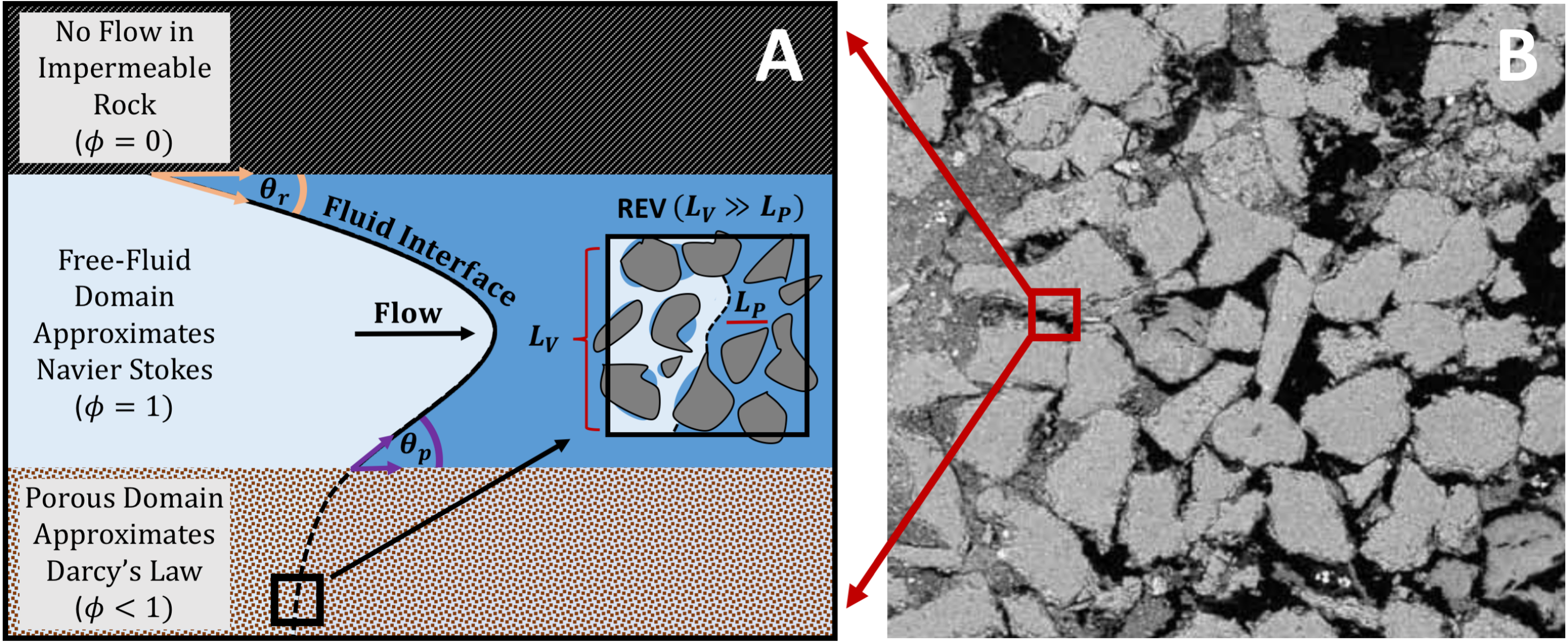}
\caption{\label{fig:concept} A) Conceptual representation of the Multiphase Micro-Continuum approach. The image shows an advancing fluid-fluid interface in a system that contains impermeable rock (top), free fluid (center), and a permeable porous medium (bottom). The two immiscible fluids are shown in different shades of blue (left and right) and the inset shows a sample REV over which the model's equations are averaged. B) SEM image of a shaly sandstone obtained from \cite{Peters2009} showing the distribution of its components: porous clay (dark grey), non-porous sand (light grey), and open pore space (black).}
\end{center}
\end{figure*}

Digital Rock Physics is made possible by advances in high resolution imaging techniques, notably X-ray Microtomography (XCT) \citep{Baker2012,Singh2018,kohanpur_valocchi_2020} and focused ion beam scanning electron microscopy (FIB-SEM) \citep{Cnudde2013,Kelly2016,Welch2017,ruspini2021}. The first method, XCT, involves recording hundreds or thousands of two-dimensional (2-D) X-ray scans through a sample that are then computationally reconstructed to create a 3-D image. This method enables detailed volumetric representations of rock core samples spanning several cubic millimeters with a resolution of about 1 cubic micrometer \citep{Wildenschild2013,Blunt2013}. The second method, FIB-SEM, involves repeated etching and imaging of a sample through alternating application of focused ion beams and scanning electron microscopy at considerably smaller scales. It yields images spanning $\sim$5 cubic micrometers with an associated resolution of $\sim$5 cubic nanometers \citep{Dewers2012}. These two techniques highlight an important limitation of current imaging techniques: the existence of an unavoidable trade-off between image resolution and field-of-view. 

The inherent complexity of most natural rocks further complicates the imaging and characterization process. More often than not, rocks such as sandstones, carbonates, and shales exhibit heterogeneities that span several length scales \citep{Bear1988,Mousavi2013,Akbarabadi2017,Beckingham2017}, some of which cannot be properly resolved by the aforementioned imaging techniques. A common way to simplify the imaging process while accounting for these heterogeneities is to designate a ``cutoff" voxel size that resolves the largest pores within a given rock sample (or other features of interest such as cracks or fractures) while simultaneously acting as a ``filter" for any pores smaller than that particular size. These small pores, which are not individually resolved, are then designated ``sub-resolution porosity" (SRP) and labeled as a third phase in the rock-pore-SRP system during the eventual segmentation of sample images. The final result is a reconstructed image with an acceptable trade-off between resolution and field of view \citep{Scheibe2015}. 

Until recently, and despite its abundance in reconstructed natural rock scans, SRP was generally assumed to have little influence on rock hydraulic properties predicted from flow simulations. Most computational models were based on the simplifying assumption that transport within the SRP is dominated by diffusion and thus contributes negligibly to fluid flow \citep{Haggerty1995,Carrera1998,Gouze2008,Shabro2011,Gjetvaj2015}. However, recent studies have shown that this assumption breaks down whenever SRP contributes significantly to the rock's percolating path by forming bridges between resolved pore spaces. In these cases, SRP can impact permeability by a factor larger than 2 even when contributing only $\sim$2\% of the total porosity \citep{Churchel1991,Tanino2012,Soulaine2016_subresolution,wu_tahmasebi_lin_munawar_cnudde_2019}. In addition, recent evidence suggests that SRP also can have important impacts on multiphase flow, as shown by observations of dramatic changes in relative permeability curves and overall flow behaviour associated with differences in SRP wetting properties in mixed-wet or strongly-wetting rocks \citep{Zou2018,Rucker2019,Fan2020,Garfi2020}.

\subsection{Multiscale Models}

A potential path towards resolving the influence of SRP on hydrologic processes is provided by sustained efforts to develop numerical techniques designed to account for this feature combined with steady advances in high-performance parallel computing \citep{Arbogast1993,Arbogast1993b,Moctezuma2004,Javadapour2009,Bauer2011,Jiang2013b}. In particular, the development of multiscale/dual-porosity pore network models (D-PNM) has allowed for relatively fast and accurate assessment of the permeability of rocks containing multiscale heterogeneity \citep{Bekri1995}. Classical PNMs rely on approximating the 3-dimensional resolved pore space through a series of ideally-shaped pore ``nodes" and ``throats" \citep{Fatt1956}. The result is a system where the relevant fluid dynamics can be readily solved through idealized equations for flow \citep{Dong2009,Joekar-Niasar2012,Jiang2012,Blunt2013,Huang2016,Suo2020}. In D-PNMs, the presence of SRP is accounted for through the implementation of an additional fine-scale pore network \citep{Ioannidis2000,Jiang2013,Prodanovic2015,sadeghnejad_gostick_2020,moslemipour_sadeghnejad_2020} or through the creation of ``micro-links" forming percolation paths between large pores \citep{Bultreys2015,xu_lin_jiang_ji_cao_2021}. Accurate definition of SRP connectivity within these networks remains a challenge \citep{zhao_shang_jin_jia_2017,petrovskyy_dijke_jiang_geiger_2020}, as, by definition, there are no discernible features from which to inform assignments of pore network topology within the SRP. 

The expansion of multiscale models into multiphase flow further complicates matters, as the effects of capillarity and wettability need to be modeled through representative relative permeability and capillary pressure models in order to obtain accurate flow representations within the SRP \citep{Carrillo2020}. For this reason, the few studies that implemented multiphase D-PNMs have relied on the assumption of quasi-static fluid displacement, an assumption valid for simulating flow at low capillary numbers \citep{Mehmani2013b,Bultreys2015,xu_lin_jiang_ji_cao_2021} and where both phases are effectively set at a given saturation. These studies have leveraged D-PNMs to study how the amount and distribution of SRP affects the relative permeability behaviour of artificial rock samples \citep{Mehmani2014} and how SRP characterization and connectivity affect the wetting properties of natural rocks \citep{Bultreys2016,song_yao_zhang_sun_yang_2021,isah_adebayo_mahmoud_babalola_el-husseiny_2020}. Unfortunately, due to the simplifying assumptions of D-PNMs outlined above, extension of these studies to dynamic systems with mixed-wet SRP or systems with viscously-dominated flow remains impossible.

The Micro-Continuum approach presents an alternative route to simulating dynamic flow processes in systems with SRP. This approach relies on locally-averaged Navier-Stokes equations that asymptotically approach Darcy's law in regions with SRP and the Navier-Stokes equations in fully resolved pores. This model has proven fairly flexible and has been used to evaluate the effects of static \citep{Knackstedt_2006,Apourvari2014,Scheibe2015,Soulaine2016_microcontinuum,Guo2018,Kang2019,Singh2019}, reactive \citep{Soulaine2017_dissolutionSinglePhase,Noiriel2021}, and deformable \citep{Carrillo2019} SRP on the permeability of heterogeneous porous media. Furthermore, through  careful consideration of capillary and viscous effects within the SRP (i.e., fluid mobility, relative permeabilities, and capillary pressures), recent investigations have successfully expanded and validated the Micro-Continuum Approach for situations involving the flow of multiple fluids in multiscale porous media \citep{Soulaine2018_twoPhaseDissolution,Carrillo2020,Carrillo2020MDBB,Carrillo2021}. In this approach, the impact of simplifying model assumptions is greatly reduced relative to the D-PNM approach at the expense of relatively high computational costs. As such, this approach allows for the simulation of dynamic multiscale systems in domain sizes that approach that of an REV. 

\subsection{Objective of this Paper}

In this study, we leverage the capabilities of the Multiphase Micro-Continuum Approach to systematically examine the influence of SRP properties (permeability, porosity, wettability) on Direct Numerical Simulation predictions of multiphase flow in a digital model of a carbonate rock. In particular, we characterize the rock's absolute permeability, relative permeability curves, residual permeabilities, and fluid breakthrough times on the $\sim$30 mm$^3$ scale of an XCT image. We hypothesise that the SRP properties outlined above have even greater impacts on multiphase flow than on single phase flow, such that their neglect or misrepresentation leads to inaccurate predictions of rock hydraulic properties. To the best of our knowledge, this is the first application of Direct Numerical Simulations to multiphase flow in rock samples containing unresolved porosity and the first computational effort to systematically examine the impacts of SRP wetting properties on the aforementioned rock flow properties.

\section{Materials and Methods}\label{sec:Methods}

\subsection{Mathematical model}\label{sec:mathematical_model}

The Multiphase Micro-Continuum framework for incompressible immiscible flow in rigid porous media consists of three volume-averaged partial differential equations. They describe the conservation and transport of fluid mass (Eqn.~\ref{eq:mass_conservation}), fluid saturation (Eqn.~\ref{eq:saturation_eq}), and fluid momentum (Eqn.~\ref{eq:momentum_conservation}). Once implemented in a suitable numerical solver, these equations are used to solve for the single-field pressure ($p$), the single-field fluid velocity ($\boldsymbol{U}$), and the wetting-fluid saturation ($\alpha_w$). A full description of the model can be found in \cite{Carrillo2020}. Here, we have:

\begin{equation}
\nabla\cdot\boldsymbol{U}=0,
\label{eq:mass_conservation}
\end{equation}

\begin{equation}
\frac{\partial \phi \alpha_w}{\partial t} + \nabla\cdot\left(\alpha_w \boldsymbol{U}\right) + \nabla\cdot\left(\phi \alpha_w\alpha_n\boldsymbol{U}_r\right)=0,
\label{eq:saturation_eq} 
\end{equation}

\begin{equation}
    \frac{1}{\phi }\left( \frac{\partial \rho \boldsymbol{U}}{\partial t} +\nabla \cdot \left( \frac{\rho}{\phi }\boldsymbol{U} \boldsymbol{U} \right)\right)=-\nabla   p 
+ \nabla\cdot\boldsymbol{S} -\mu k^{-1} \boldsymbol{U} + \boldsymbol{F}_c, 
\label{eq:momentum_conservation}
\end{equation}

\noindent where the subscripts $w$ and $n$ refer to the wetting and non-wetting fluids, $\phi$ is the cell porosity, $\rho$ is the single-field density, $\mu k^{-1}$ is the drag coefficient of the unresolved porous media (a function of the cell permeability, saturation, and fluid viscosities), $\boldsymbol{F}_c$ are the capillary forces, and $\boldsymbol{S}={\mu}(\nabla\boldsymbol{U}+{(\nabla\boldsymbol{U})}^T)$ is the averaged single-field shear stress tensor. Here, gravity is neglected and the phrase ``single-field" refers to averaged variables that depend on the properties of both fluids \citep{Maes2020}. 

A key feature of Eqns. \ref{eq:mass_conservation}-\ref{eq:momentum_conservation} is that they are valid in control volumes that contain any combination of the three relevant phases (porous solid, wetting fluid, non-wetting fluid), meaning that they can be applied to systems that contain both solid-free ($\phi =1$) and porous regions ($\phi <1$). Due to the scale separation hypothesis \citep{Whitaker1986}, this unique set of equations tends towards distinct solutions in solid-free and porous regions. Notably, the single-field momentum equation tends to a solution that can be asymptotically matched to the two-phase Navier-Stokes equations in solid-free regions and to two-phase Darcy's law in porous regions \citep{Carrillo2020}:
\begin{equation} \label{eq:SolidFree_Darcy}
    \begin{cases}
        \frac{\partial {\rho }{\boldsymbol{U}}}{\partial t}+\nabla\cdot\left({\rho }{{\boldsymbol{U}} \boldsymbol{U}}\right)=-\nabla p+\nabla\cdot\boldsymbol{S}+{\boldsymbol{F}}_c, & \textnormal{if } \phi = 1, \\
        {\boldsymbol{U}}=-\frac{k}{\mu }\left(\nabla p-{\boldsymbol{F}}_c\right), & \textnormal{if }  \phi < 1.
    \end{cases}
\end{equation}
As such, the Multiphase Micro-Continuum model is ideally suited for simulating multiphase flow in XCT images that contain SRP, as illustrated schematically in Figure~\ref{fig:concept}. 

The asymptotic matching noted above requires appropriate definitions of the relative velocity $\boldsymbol{U}_r$, drag force $\mu k^{-1} \boldsymbol{U}$, and capillary forces $\boldsymbol{F}_c$. These variables reflect the influence of sub-grid-scale structure and dynamics, including the fluid distribution and the impact of porous micro-structure on flow within the SRP. For this reason, these parameters are defined differently in the solid-free region ($\phi =1$) and porous regions ($\phi <1$). In particular, the single-field drag force is negligible in solid-free regions and, in porous regions, depends on absolute ($k_0$) and relative permeabilities ($k_{r,i}$) within the SRP:
\begin{equation} \label{perm}
\mu k^{-1}=
    \begin{cases}
        0, & \textnormal{if } \phi = 1,\\
        k_0^{-1}\left(\frac{k_{r,w}}{\mu_{w}}+\frac{k_{r,n}}{\mu_{n}}\right)^{-1}, & \textnormal{if }  \phi < 1.
    \end{cases}
\end{equation}

The capillary forces within the solid-free region are proportional to the surface tension $\gamma$ and the curvature of the fluid-fluid interface as described by the Continuum Surface Force formulation \citep{Brackbill1992}. In the porous region, capillary forces are a function of the fluid mobilities ($M_i=k_0k_{i,r}/{\mu }_i; \ M=M_w+M_n$) and the average capillary pressure $p_c$: 
\begin{equation} 
\boldsymbol{F}_{c}=\begin{cases}
-\gamma\nabla.\left(\hat{\boldsymbol{n}}_{w,n}\right)\nabla \alpha_{w}, & \textnormal{if } \phi = 1,\\
\left[M^{-1}\left( M_w \alpha_n - M_n \alpha_w \right)\left(\frac{\partial p_c}{\partial \alpha_w }\right) - p_c \right]\nabla \alpha_w, &  \phi < 1,
\end{cases}
\label{eq:surface_tension_forces}
\end{equation}

\noindent where the normal at the fluid-fluid interface, $\hat{\boldsymbol{n}}_{w,n}$, is given by
\begin{equation}\label{normal}
\hat{\boldsymbol{n}}_{w,n}=\begin{cases}
-\frac{\nabla\alpha_w}{\left|\nabla\alpha_w\right|}, & \textnormal{if } \phi = 1,\\
\cos \theta_p \boldsymbol{n}_{wall} + \sin \theta_p \boldsymbol{t}_{wall}, &  \textnormal{at the SRP surface}.
\end{cases}
\end{equation}

Equation \ref{normal} imposes a contact angle $\theta_p$ at the SRP surface following the approach developed by \cite{Horgue2014}, where $\boldsymbol{t}_{wall}$ and $\boldsymbol{n}_{wall}$ are the tangential and normal directions relative to the SRP surface. The specification of the contact angle at non-porous rock surfaces, $\theta_r$, follows a similar implementation.

The relative fluid velocity is given by:

\begin{equation}
\boldsymbol{U}_{r}=\begin{cases}\label{Ur}
C_\alpha \max \left(\left| \boldsymbol{U} \right|\right) \frac{\nabla \alpha_w}{\left| \nabla \alpha_w \right|}, &  \textnormal{if } \phi = 1, \\
 {\phi}^{-1}\left[ \begin{array}{c}
 -\left(\frac{M_w}{\alpha_w} - \frac{M_n}{\alpha_n}\right)\nabla  p  \\ +\left(\frac{M_w\alpha_n}{\alpha_w} +\frac{M_n\alpha_w}{\alpha_n} \right)\nabla p_c \\ - \left(\frac{M_w}{\alpha_w} - \frac{M_n}{\alpha_n} \right)p_c \nabla \alpha_w \end{array}\right], &  \textnormal{if } \phi < 1,
\end{cases}
\end{equation}

\noindent where $C_{\alpha}$ is an interface compression parameter used in the Volume-of-Fluid method (typically set to values between 1 and 4), and the expression within the SRP is imposed by asymptotic matching to two-phase Darcy's law \citep{Carrillo2020}.

Lastly, closure of the system of equations presented above requires appropriate constitutive models to solve for $p_c$ and $k_{r,i}$ within the SRP. For simplicity, we use the well-known Van Genuchten model \citep{VanGenutchen1980}:

\begin{equation}
k_{r,n} = (1-\alpha_{w})^{1/2}(1-\alpha_{w}^{1/m})^{2m},
\label{eq:rel_perm}
\end{equation}

\begin{equation}
k_{r,w}= \alpha_{w}^{1/2}(1-(1-\alpha_{w}^{1/m})^m)^{2},
\end{equation}

\begin{equation}
    p_c\ ={p_{c,0}\left({\left({\alpha }_{w}\right)}^{-\frac{1}{m}}-1\right)}^{1-m}, 
\label{eq:p_cap}
\end{equation}

\noindent where $m$ is a wetting parameter that controls the internal wettability of the SRP and $p_{c,0}$ is the entry capillary pressure of the SRP. The SRP is internally water-wet if $m > 1$ and oil-wet if $m <1$. Note that the sign of the entry capillary pressure was changed for values of $m > 1$ to prevent unphysical parameterizations where the SRP is both water-wet and oil-wet at the same time. 

\subsection{Numerical Implementation}

The mathematical model presented in Section \ref{sec:mathematical_model} was numerically implemented in OpenFOAM{\circledR}, a free, parallel, C++ simulation platform that uses the Finite Volume Method to discretize and solve partial differential equations in three-dimensional grids. Mass conservation and incompressibility (Eqns.~\ref{eq:mass_conservation} and \ref{eq:momentum_conservation}) were ensured through the Pressure Implicit Splitting-Operator (PISO) algorithm \citep{Issa1986}. The evolution of the fluid-fluid interface (Eqn.~\ref{eq:saturation_eq}) was solved using the Multidimensional Universal Limiter of Explicit Solution (MULES) algorithm \citep{Marquez2013} and a Piecewise-Linear Interface Calculation (PLIC) compression scheme. Extensive validation of the modeling framework is presented in \cite{Carrillo2020} and the open-source implementation is available from the author's \href{https://github.com/Franjcf}{GitHub} repository \citep{hybridPorousInterFoam_code}.

\subsection{Studied Rock Sample} \label{sec:sample}

\begin{figure}[t!]
\begin{center}
\includegraphics[width=0.48\textwidth]{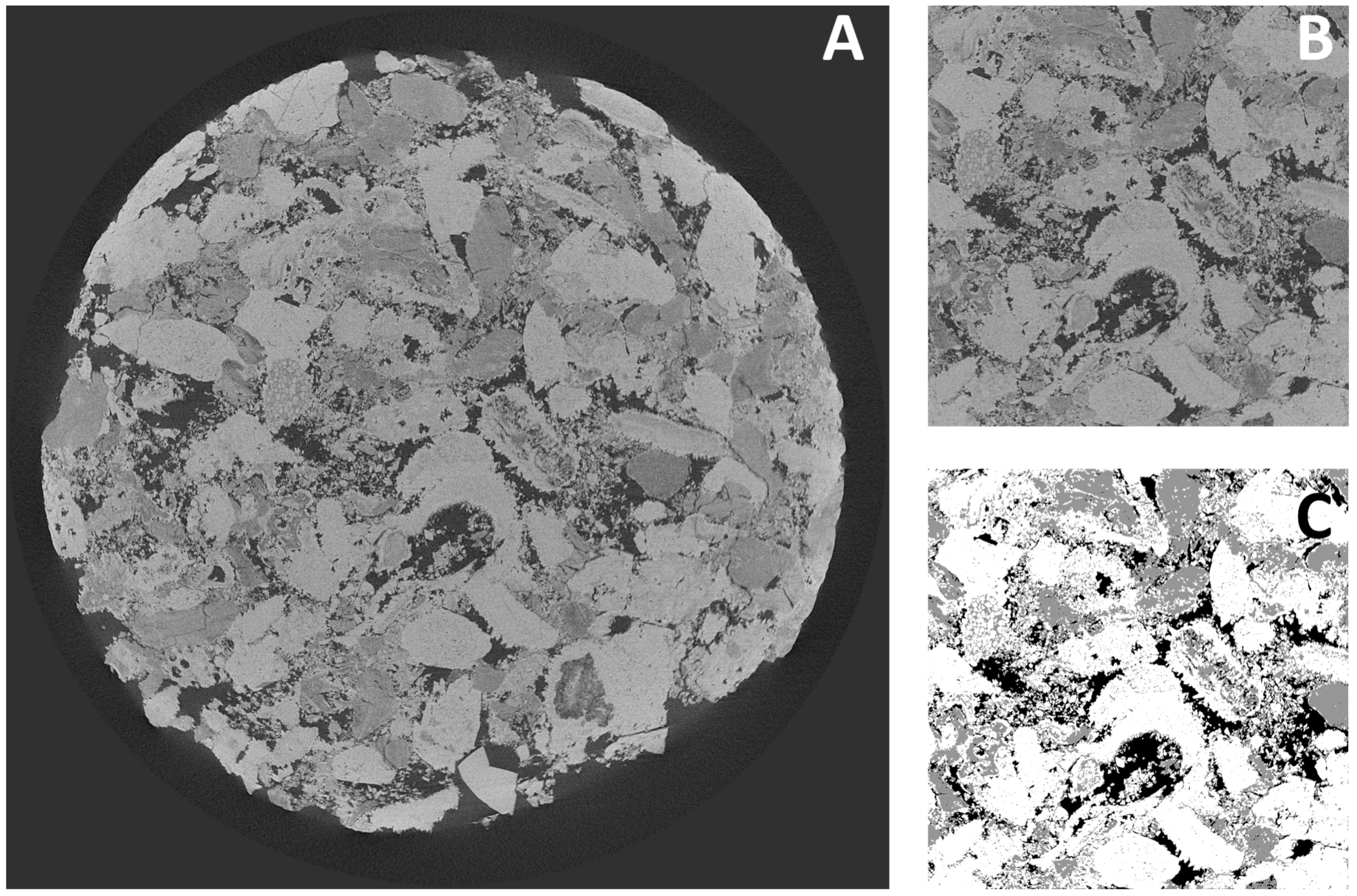}
\caption{\label{fig:sample} Representative cross-section of the XCT scan of Estaillades carbonate rock used in this study \citep{sample_dataset}. A) Full 2-D view of the sample, which is 7 mm in diameter with a resolution of 3.1 $\mu \mathrm{m}$ per voxel. B) 500 by 500 voxel cropped sample. (C) Corresponding segmented image. For all figures, black is open pore space, dark grey corresponds to domains that contain SRP, and the lightest color is solid calcite.}
\end{center}
\end{figure}

Simulations were performed on a reconstructed 3-D XCT scan of an Estaillades Carbonate sample obtained from \cite{sample_dataset} through the \href{https://www.digitalrocksportal.org/}{Digital Rock Portal}. This set of images has been used in several previous D-PNM studies \citep{Bultreys2015,Bultreys2016}. The sample (1000 by 1000 by 1000 voxels, $3.1 \ \mu \mathrm{m}$ per voxel) is ideally suited for our purposes, as it is a mono-mineralic calcite rock containing both intergranular macropores and unresolved intragranular micropores (i.e., SRP). Voxels containing solid rock, resolved pores, and unresolved pores were identified through a 3-phase segmentation procedure following the steps outlined in \citet{Bultreys2015}. This yielded a sample with $56.2 \%$ solid rock voxels, $11.8\%$ porous voxels, and $32\%$ microporous voxels (Figure \ref{fig:sample}). 

\begin{figure}
\begin{center}
\includegraphics[width=0.48\textwidth]{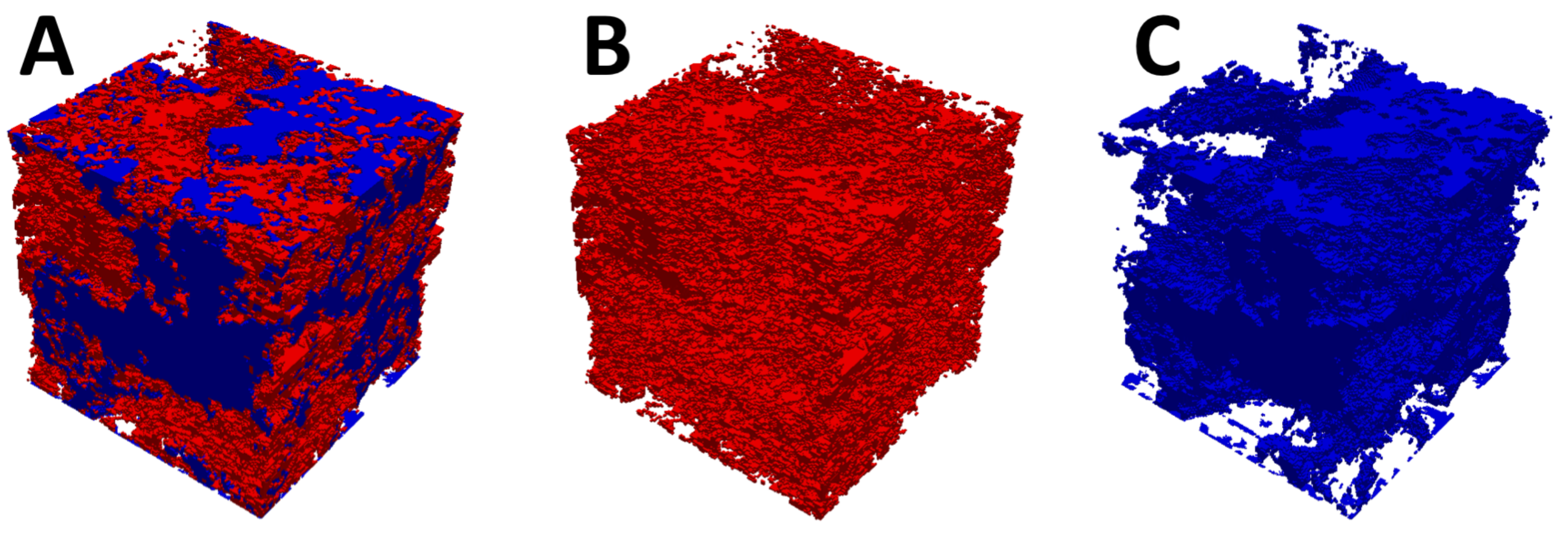}
\caption{\label{fig:porosity_distribution} Spatial distribution of the SRP (red), pore space (blue), and solid rock (transparent) within the extracted 3-D rock representation (200 by 200 by 200 cells). A) The complete computational mesh. B) The corresponding SRP, which accounts for $21\%$  of the voxels. C) The associated open pore space, which accounts for $40 \%$ of the voxels.}
\end{center}
\end{figure}

Due to the computational cost associated with performing direct numerical simulations on such a large physical space, we extracted a 200 by 200 by 200 voxel sub-sample from the original scan in order to perform our simulations. The computational cost was further reduced by removing all grid cells corresponding to solid rock voxels in the resulting computational mesh, yielding a sample of about 3.2 million cells (see Fig. \ref{fig:porosity_distribution}). In order to maintain adequate mesh resolution while properly representing the mobile fluid-fluid interface within the open pore space, we implemented a dynamic mesh refinement algorithm that allowed the mesh to become up to 16 times finer at said interface. No mesh refinement was carried out within the SRP. Lastly, as is customary for these types of simulations and to properly control the flow rate into the sample, we added two ``buffer" regions at the inlet and outlet boundaries of our samples. All other boundaries were defined with no-flow boundary conditions. 

\section{Base Simulation Setup and Upscaling}\label{sec:simulation_setup}

\subsection{Base Simulation Parameterization}\label{sec:base_simulation_parameterization}

The main purpose of our simulations is to perform a sensitivity analysis of the impact of SRP properties on single and multiphase flow at the scale of the full digital rock image, hereafter referred to as the macroscopic scale. For this, we first introduce a ``base" simulation that will be parameterized using experimental values and then used as a template for the systematic variation of SRP properties. This workflow is conceptually similar to the one performed in \cite{hashemi_blunt_hajibeygi_2021}.

Our base simulation involves the injection of oil into a fully-water-saturated rock sample at a constant rate of 0.1 $\mu$L s$^{-1}$ until the simulation reaches a steady-state. The choice of the labels ``oil" and ``water" is an arbitrary one: our main goal is to examine the flow of two immiscible and incompressible fluids. The advancing fluid is non-wetting in our base simulation, but the wettability of the solid by the two fluids is reversed in some of our simulations. The rock and fluid properties are summarized in Tables \ref{table:fluid_parameters} and \ref{table:rock_parameters}. The subsequent sensitivity analysis was performed by independently modifying the SRP's porosity ($\phi = 0$ to $1$), absolute permeability ($k_0 = 10^{-12}$ to $10^{-17} \ m^2$), internal wetting properties ($m = 0.2$ to $1.5$ in Eqns.~\ref{eq:rel_perm}-\ref{eq:p_cap}), and the contact angles formed by fluid-fluid interfaces on the external surface of SRP and impermeable rock domains ($\theta_p$ and $\theta_r = 30^{\circ}$ to $150^{\circ}$). 
The decoupling of the contact angle at the SRP and impermeable rock surfaces allows the investigation of mixed-wet systems \citep{Song2015,Huang2016,Akbarabadi2017} and establishes the possibility of defining a roughness- or saturation-dependent contact angle in future studies \citep{Wenzel1936,Whyman2008}. The decoupling of internal ($m$) and surface ($\theta$) wetting properties allows us to differentiate between macroscopic and microscopic wetting effects, where $\theta$ impacts multiphase flow in large pores and $m$ impacts multiphase flow within the SRP. 

Lastly, we carried out additional single-phase flow simulations for each case where we varied the SRP porosity and permeability. This was necessary to calculate each case's absolute permeability and relative permeability curves (see Section \ref{sec:relative_perm_calc}). On average, each multiphase simulation ran for approximately 120 hours on ten 28-core Broadwell Xeon nodes.




\begin{table}[!htb]
    \centering
    \setcellgapes{2pt}
    \makegapedcells
    \medskip
\begin{tabular}{||c  c||} 
\hline
 Property & Value \\ [0.5ex] 
 \hline\hline
 $\rho_w$ & \SI{1000}{kg.m^{-3}} \\
 \hline
 $\mu_w$ &\SI{0.001}{Pa.s}\\
 \hline
 $\rho_n$ & \SI{800}{kg.m^{-3}}  \\
 \hline
 $\mu_n$ & \SI{0.1}{Pa.s} \\
  \hline
 $\gamma$ & \SI{0.03}{kg.s^{-2}} \\
  \hline
\end{tabular}
 \caption{Simulated Fluid Properties. These were kept constant for all simulations.}
    \label{table:fluid_parameters}
\end{table}

\begin{table}[!htb]
 \centering 
 \setcellgapes{2pt}
 \makegapedcells
 \medskip
 \begin{tabular}{||c c c||} 
 \hline
 Property & Base Value & Range \\ [0.5ex] 
 \hline\hline
 \(\phi\) & 0.5 & $0 - 1$ \\ 
 \hline
 \(k_0 \) & \SI{e-13}{m^{2}} & $10^{-12} -$ \SI{e-17}{m^{2}} \\ 
 \hline
 \(\theta_r \) & $30^{\circ}$ & $30 - 150^{\circ}$  \\
 \hline
 \(\theta_p \) & $30^{\circ}$ & $30 - 150^{\circ}$   \\
 \hline
 $m$ & 1 & $0.2-1.5$ \\ 
 \hline
  \(p_{c,0}\) &  $\pm$ \SI{1.35e4}{Pa} & n/a \\
 \hline
\end{tabular}
 \caption{Simulated Rock and SRP Parameters. The second column represents each parameter value used in the base simulation and the third column shows the range over which each parameter was varied. These ranges where chosen to create representative samples of the associated parameter space: from strongly hydrophobic to strongly hydrophilic systems, and from systems with permeable (and impermeable) SRP to systems with no SRP.}
    \label{table:rock_parameters}
\end{table}

\subsection{Calculation of Absolute Permeability and Relative Permeability Curves}\label{sec:relative_perm_calc}

Relative permeabilities were calculated through modification of the upscaling approach presented in \cite{Raeini2014}, where macroscopic relative permeability $K_{r,i}$ is defined as the ratio between the apparent permeability $K_i$ calculated from transient, multi-phase flow experiments and the upscaled absolute permeability $K_0$ calculated from steady-state, single-phase flow experiments: 

\begin{equation}\label{eq:Kr}
    K_{r,i} =  \frac{K_i}{K_0} =  \frac{Q_i/\Delta P_i}{Q_{i,s}/\Delta P_{i,s}}.
\end{equation}

In Equation \ref{eq:Kr}, the subscript $i$ identifies properties pertaining to either fluid and the subscript $s$ refers to quantities obtained from single-phase experiments. Furthermore, $Q_i = \int \boldsymbol{U}\cdot \boldsymbol{n}\alpha_i dA$ is the volumetric fluid flow rate of phase $i$ passing through an area $A$ into the porous medium, and $\Delta P_i$ is the pressure drop in phase $i$ across said medium. The latter is defined as follows: 


\begin{align}\label{eq:deltaP}
    \Delta P_{i} \equiv & -\frac{1}{Q_{i}} \int_{V_f}\left( -\nabla p   + \boldsymbol{F}_c \right)\cdot \boldsymbol{U}d{V_{f,i}},
    \\ 
    & = -\frac{1}{Q_{i}}\int_{V_{f}} \nonumber \left(\frac{\mathrm{D}}{\mathrm{D}t}(\rho\boldsymbol{U}) -  \nabla\cdot\boldsymbol{S} + \mu k^{-1} \boldsymbol{U}\right) \cdot \boldsymbol{U}d{V_{f,i}},
\end{align}

\noindent where $V_f$ is the fluid volume of the sample excluding the buffer zones. A drag term ($\mu k^{-1} \boldsymbol{U}$) is included Equation \ref{eq:deltaP} to account for the momentum dissipation (i.e. pressure drop) induced by the presence of SRP in the sample. The calculation of $\Delta P_{i,s}$ follows Eqn. \ref{eq:deltaP} sans the capillary force term. 

Relative permeability curves were constructed by matching each $K_{r,i}$ value to the corresponding saturation in the porous medium at a specific point in time. This so-called ``unsteady" approach, where $K_{r,i}$ values are not calculated at steady state \citep{Amaefule1982,Johnson1959}, enables calculating relative permeability curves without needing to carry out a distinct steady-state multiphase simulation for each data point, a current necessity across numerical frameworks for rock models with realistically complex pore structures given current computational capabilities. However, this comes at the expense of accuracy or, more precisely, at a risk that the resulting relative permeability curves may be sensitive to fluid flow rate \citep{Diamantopoulos2012}. To minimize the impact of this approximation, we focus on characterizing the sensitivity of $K_{r,i}$ to different SRP properties, as opposed to absolute values of $K_{r,i}$. In other words, we aim to gain insight into the magnitude of the different impacts of SRP on multiphase flow, as opposed to quantitatively matching experimental results. 

\section{Impact of SRP on Absolute Permeability}\label{sec:SRP_effects_on_absolute_perm}

In the following three sections, we quantify the effects of SRP properties on the rock's overall absolute permeability (this section), relative permeability curves (Section \ref{sec: SRP_effects_on_rel_perm}), and time-dependent saturation profiles (Section \ref{sec:SRP_Effects_On_Saturation}). For the remainder of this study, each simulation case is identified by the variable that is changed with respect to the base simulation established in Section \ref{sec:base_simulation_parameterization} and parameterized according to Tables \ref{table:fluid_parameters} and \ref{table:rock_parameters}. We now start by evaluating the effect of SRP on the rock's absolute permeability.

\begin{figure}[h!]
\includegraphics[width=0.48\textwidth]{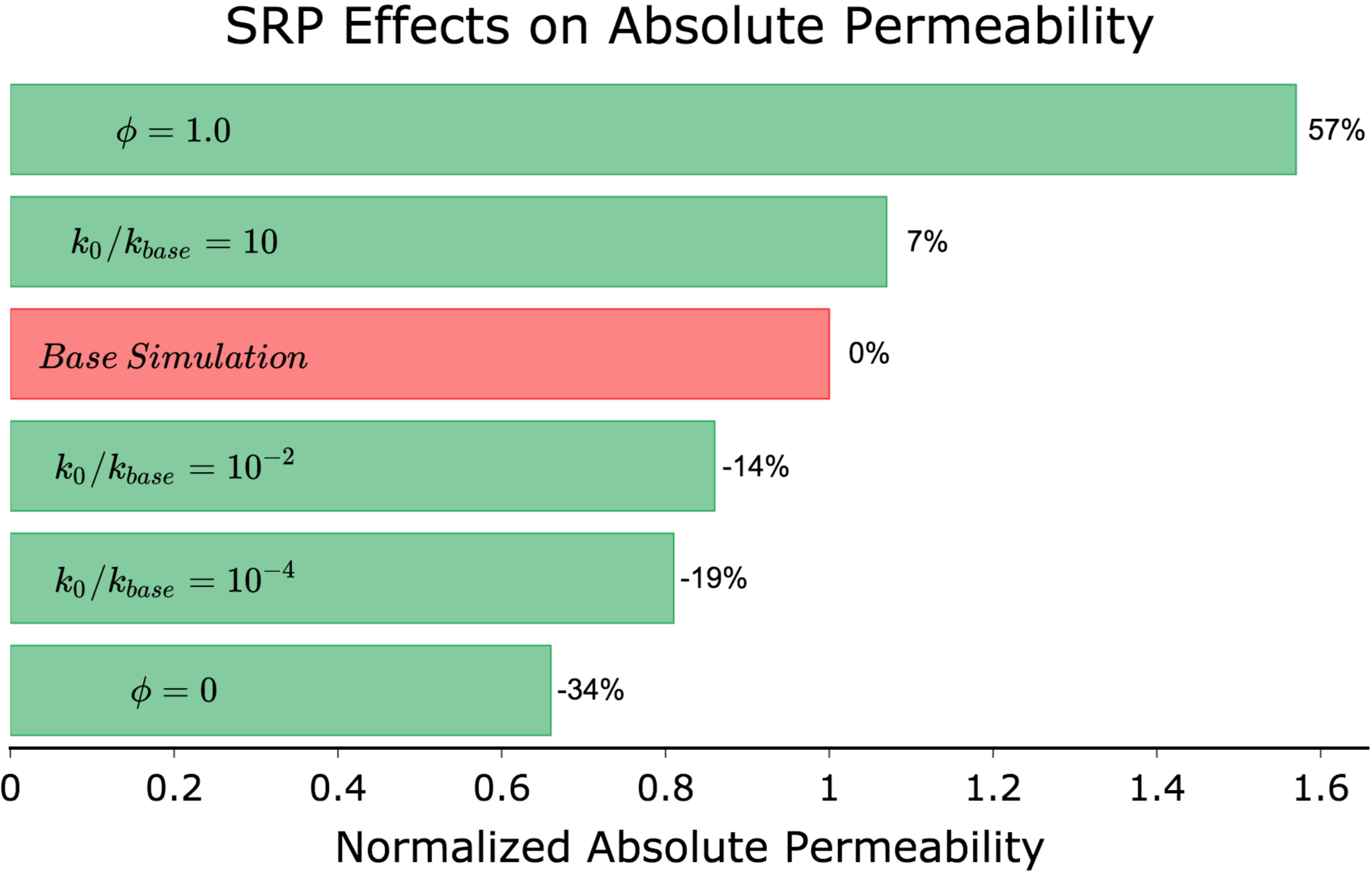}
\caption{Sample absolute permeability as a function of SRP properties. Each label shows the only varied parameter with respect to the base simulation. Values indicated to the right of each bar show the percent change in absolute permeability relative to the base simulation described in Section \ref{sec:base_simulation_parameterization}.\label{fig:abs_perm}}
\end{figure}

Figure \ref{fig:abs_perm} shows that the sample's absolute permeability is overestimated by $57\%$ if the SRP is neglected and assumed to be open pore space ($\phi = 1$) and underestimated by $34\%$ if it is ignored and assumed to be impermeable ($\phi = 0$), where the former's permeability more than doubles the latter. The overall trend in Figure \ref{fig:abs_perm} is fairly intuitive: as the SRP's porosity and/or permeability increase, so does the rock's absolute permeability. This is in line with the findings of \citet{Mehmani2014}, and \citet{Soulaine2016_subresolution}. However, whereas some previous studies have observed that SRP can have a disproportionately large impact on permeability, implying that it forms key percolation pathways for single-phase flow \citep{Soulaine2016_subresolution}, the factor of $\sim$2 impact of SRP on absolute permeability observed here is roughly consistent with the predictions of the well-known Kozeny-Carman equation, implying that SRP is relatively uniformly distributed in the studied rock sample (in close agreement with \citet{Bultreys2016}). As examined in the following sections, greater impacts of SRP are observed in systems with multiple fluid phases, where SRP wettability and relative permeability become key factors controlling fluid flow.


\section{Impact of SRP on Relative Permeability Curves} \label{sec: SRP_effects_on_rel_perm}

Changes in sample relative permeability as a function of SRP porosity, wetting properties, and absolute permeability are not particularly intuitive. These often involve non-linear behaviors brought about by the combination of capillary forces and the sample's geometry. Throughout the following discussion we will see that the SRP has two primarily competing effects: it \textit{enhances} flow by connecting otherwise-isolated macroscopic flow paths, but it \textit{reduces} flow by being less permeable than the open pore space. We will show that the balance between these two roles is strongly dictated by SRP properties.

\begin{figure*}[h!]
\includegraphics[width=1\textwidth]{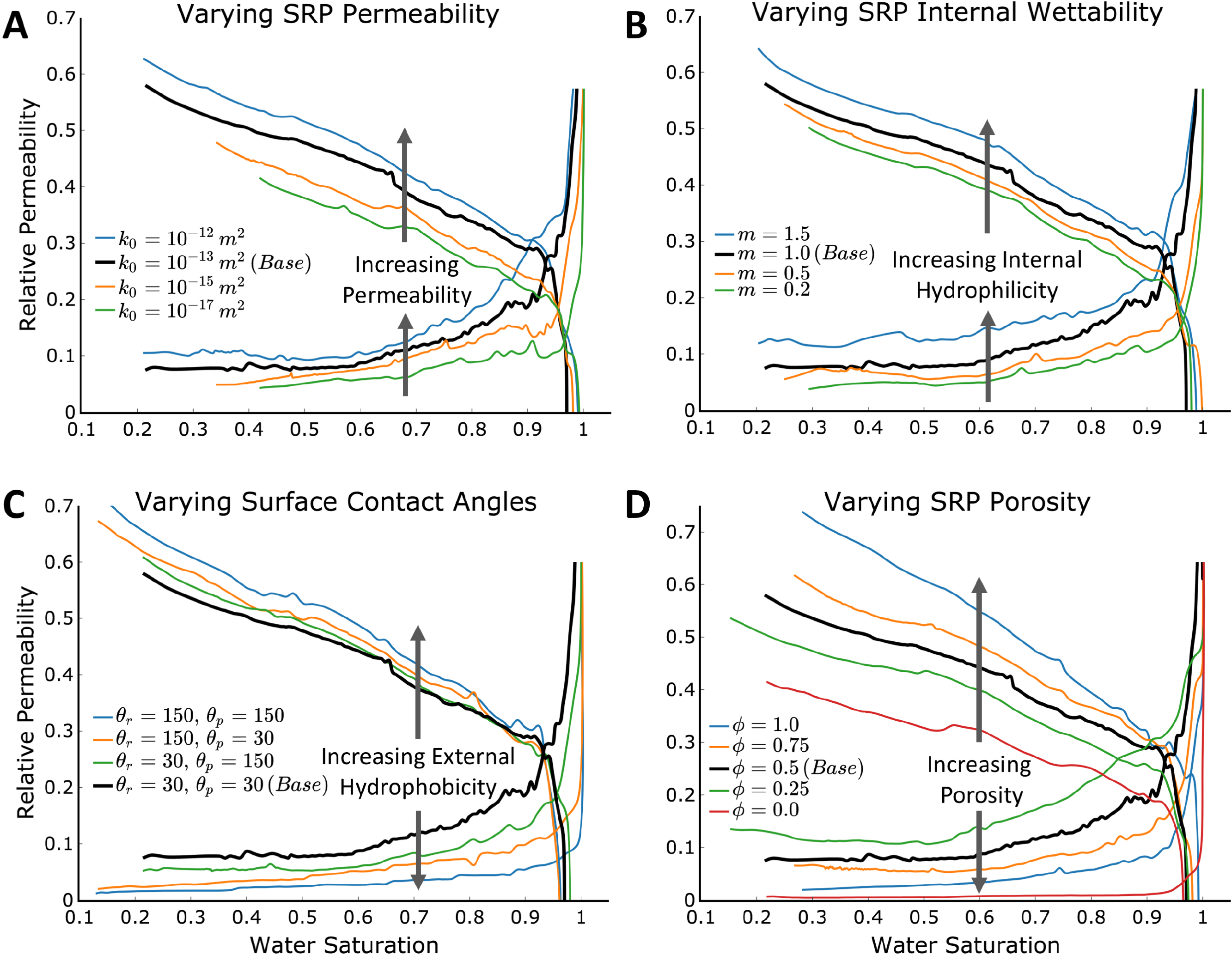}
\caption{\label{fig:all_curves} Sensitivity of drainage and imbibition relative permeability curves to different SRP properties. A) Sensitivity to SRP absolute permeability, from $k_0 = 10^{-17}$ to $10^{-12}$ m$^2$. B) Sensitivity to SRP internal wettability, from oil-wetting ($m<1$) to water-wetting ($m>1$). C) Sensitivity to the external wettability of rock and SRP domains, from water-wetting ($\theta_r = 30^{\circ} , \ \theta_p = 30^{\circ} $), to mixed-wetting ($\theta_r = 150, \ \theta_p = 30^{\circ} $ and $\theta_r = 30^{\circ} , \ \theta_p = 150^{\circ} $), to oil-wetting ($\theta_r = 150^{\circ} , \ \theta_p = 150^{\circ} $). D) Sensitivity to SRP porosity, from $\phi = 0$ to $\phi = 1$. Unless specified, all parameterized values not indicated in the legend are held constant and equal to the values described in Tables \ref{table:fluid_parameters} and \ref{table:rock_parameters}. Each color pair represents the oil (top) and water (bottom) relative permeability curves for a given simulated case. The base simulation is shown in black for all cases.}
\end{figure*}

The four sets of relative permeability curves present in Figure \ref{fig:all_curves} exhibit two distinct behaviors reflecting different responses to changes in SRP properties. In one observed behavior, the curves for both fluids shift up (or down) \textit{in the same direction} with respect to the y-axis. This implies that the sample becomes more (or less) permeable to \textit{both} phases simultaneously. In the other observed behavior, the water and oil relative permeability curves shift up or down \textit{in opposite directions}, indicating that an increased permeability to one fluid is associated with a decreased permeability to the other fluid.

\subsection{Sensitivity to SRP Absolute Permeability}\label{sec:SRP_absolute_perm}

Figure \ref{fig:all_curves}A demonstrates that an increase in SRP absolute permeability enhances the relative permeability curves of \textit{both} oil and water. This enhancement occurs in addition to the enhancement in absolute permeability presented in Fig. \ref{fig:abs_perm}. The enhancement of water relative permeability is entirely expected as the SRP is water-wet (and mostly water-saturated) in this scenario, such that greater SRP permeability naturally facilitates the flow of water. The enhancement of oil permeability is less intuitive. Since oil minimally accesses the SRP in this scenario, this enhancement is likely indirect, i.e., greater SRP permeability facilitates water drainage from the open pore space, which in turn facilitates the flow of oil.

We note, that both effects essentially disappear at SRP permeabilities below $\sim$10$^{-17}$ m$^2$ as shown in Fig. SI1 in the Supporting Information. In short, SRP permeability is only important if it is sufficiently large that flow can actually occur within the SRP.

\subsection{Sensitivity to SRP's internal wettability}\label{sec:SRP_internal_wetting_properties}

Figure \ref{fig:all_curves}B shows that an increase in SRP internal wettability, from oil-wetting ($m<1$) to water-wetting ($m>1$), also enhances the flow potential of both fluids. This effect is likely analogous to that observed for SRP absolute permeability: a more hydrophilic SRP should remain more fully water-saturated, and hence more permeable to water (because of impact of saturation on relative permeability within the SRP). As in Fig. \ref{fig:all_curves}A, this greater ability of water to flow through the SRP indirectly facilitates oil flow, likely by aiding water drainage from the open pore space.


\subsection{Sensitivity to SRP and rock surface contact angles}\label{sec:SRP_wetting_properties}

Figure \ref{fig:all_curves}C shows that the relative permeability curves shift in \textit{opposite} directions in response to changes in the external wettability of the rock or SRP surfaces. Specifically, as the pore walls become more hydrophobic, permeability to water decreases, while permeability to oil increases. The impact on oil flow is relatively small, likely because of the partial cancellation of two competing effects: more hydrophobic surfaces should inhibit oil flow by causing this flow to occur preferentially in smaller pores or closer to the pore walls; simultaneously more hydrophobic surfaces should enhance oil flow by minimizing the tendency towards trapping of oil droplets through capillary effects. Therefore, we posit that a decrease in capillary number (Ca) or sample homogeneity would likely enhance the trapping effect and may reverse the order of the oil relative permeability curves.

The impact on water flow is larger, a counter-intuitive observation. If water flows predominantly within the SRP, the impact of surface contact angles on water flow should be minimal. Alternatively, if water flows predominantly in the open pore space, surface contact angles should have relatively minor impact on relative permeability to water because of the competing effects noted above in the case of oil. In fact, an increase in water relative permeability with $\theta$ (opposite to that observed here) was reported by \cite{Fan2020}. A possible explanation of our results is that residual water flow in our simulated system relies on the \textit{combination} of SRP and residual macropore water flow (previous studies have largely ignored the presence of SRP). In systems with no microporosity, water can be retained in the open pore space through capillary forces, such as in capillary film coatings on rough pore walls \citep{Tokunaga1997,Khishvand2016}. Hydrophobic microporous walls would eliminate the capillary macropore water component of these residual flow paths.

\subsection{Sensitivity to SRP Porosity}

The effects of modulating SRP internal porosity between 0 to 1 are shown in Figure \ref{fig:all_curves}D. The overall magnitude of the relative permeability changes is in close agreement with \cite{Mehmani2014}, where the authors found that the addition of pore-clogging SRP can modify the relative permeability of the wetting and non-wetting phases by about a factor of 2. We note, again, that this effect occurs in addition to the significant impact of SRP porosity on absolute permeability presented in Fig. \ref{fig:abs_perm}.

In addition to this significant influence of SRP porosity on relative permeability, our results also show unexpected complexity. In particular, the impact of SRP porosity on water flow is non-monotonous, with minimum water relative permeabilities observed at either $\phi = 0$ or 1 and larger water relative permeabilities observed at intermediate $\phi$ values. This observation is consistent with the expected trend if residual water flow relies on a combination of both SRP and residual macropore water as suggested above: values $\phi = 0$ or 1 would inhibit water flow by eliminating the SRP water component of these residual flow paths.

\section{Impact of SRP on Residual Relative Permeability}

As noted above, our results strongly suggest that the SRP can function as an efficient and persistent connector between otherwise-disconnected water bodies, particularly at low water saturations. We call this increase in permeability the `SRP-enhanced relative permeability'. A key manifestation of this is the persistence of significant relative permeability in the water phase at water saturations below 0.5, in agreement with experimental observations for rocks with significant microporosity \citep{Bennion2010}. In contrast, pore network model simulations of multiphase flow generally predict that relative permeability to water is nearly zero at water saturations below $\sim$0.2 to 0.5 \citep{Prodanovic2015,Huang2016}.

\begin{figure}[h!]
\includegraphics[width=0.48\textwidth]{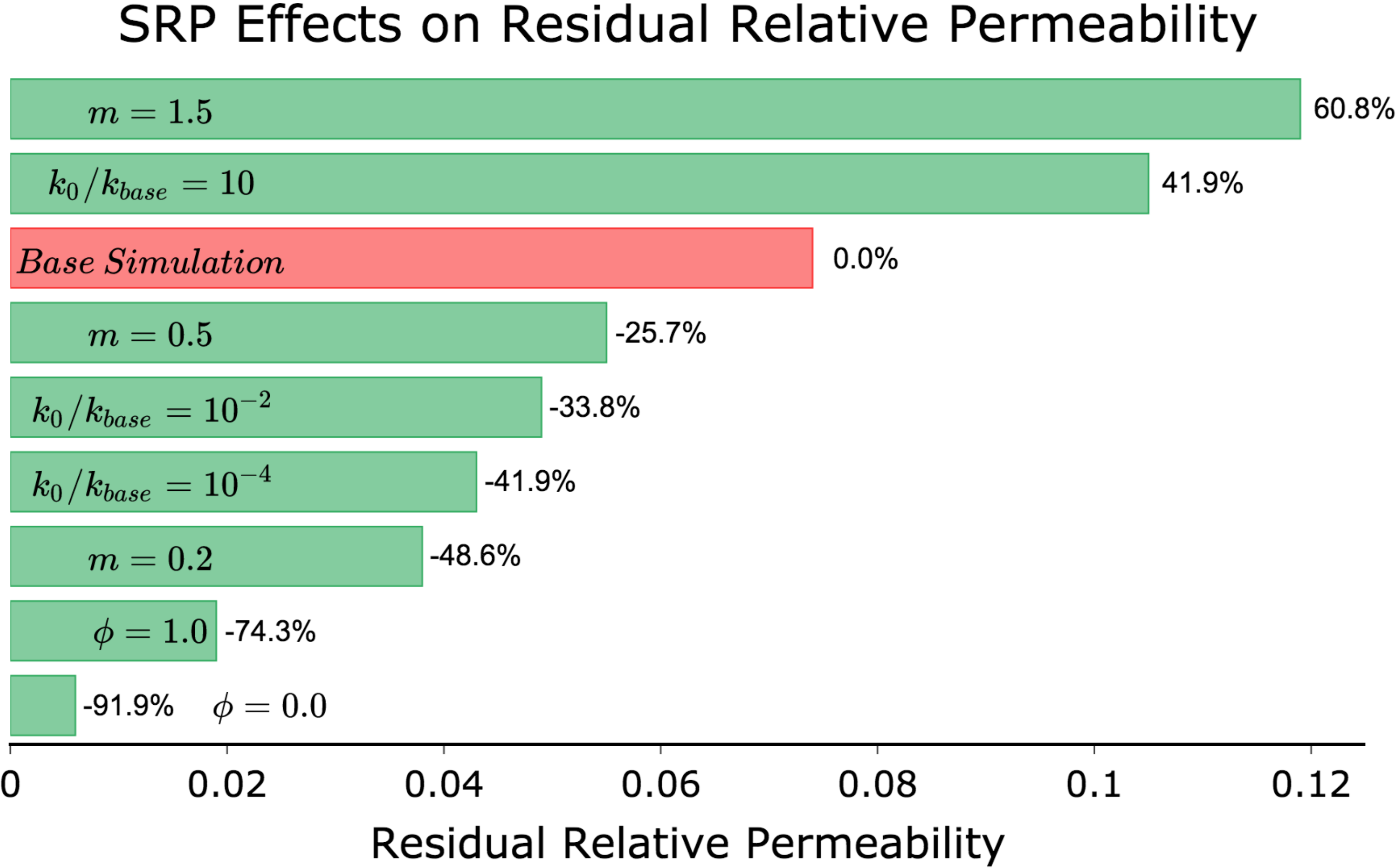}
\caption{\label{fig:residual_relative_perm} Steady-state water relative permeability for select cases. Each label shows the only varied parameter with respect to the base simulation. The percentages to the right of the bars show the percent change in residual permeability with respect to the aforementioned base simulation specified in Section \ref{sec:base_simulation_parameterization}.}
\end{figure}

A convenient way to characterize this effect is by ranking the relative permeabilities of water once each system has achieved a steady state, as seen in Figure \ref{fig:residual_relative_perm}. The overarching trend is clear: increasing the SRP's permeability to water also increases the steady-state relative permeability of said fluid (up to 20 times). The reason for this is not obvious, higher SRP permeability should lead to higher displacement of the defending fluid, lower residual saturations, and thus, lower (not higher!) steady-state permeabilities. This leads us to believe that increasing the flow capability of the SRP also leads to the creation of enhanced percolation pathways that are persistent and remain connected throughout the sample, even at low water saturations. This phenomenon is consistent with experiments in mixed-wet porous media \citep{AlRatrout2018} and somewhat analogous to thin-film flow in soils, where small amounts of water facilitate transport above the soil's water table \citep{Tokunaga1997,Lebeau2010}. This persistence of significant residual relative permeability to water has potentially important implications in the physics of soil drying \citep{Or2013} and in hydrocarbon recovery from tight sandstone formations \citep{Tian2019}.

\section{Impact of SRP on Dynamic Saturation Evolution}\label{sec:SRP_Effects_On_Saturation}

The presence of SRP has the following competing effects on the evolution of oil saturation within the sample during oil-flooding: 1) It increases the residual saturation of its wetting phase (be it oil or water) by acting as a fluid reservoir that ``defends" itself against the non-wetting phase. 2) It decreases the residual saturation of the defending fluid phase by adding additional inter-pore connectivity and outflow routes \citep{Mehmani2014}. The balance between these two effects is dictated by the flow properties of the SRP. 

\begin{figure*}[ht!]
\begin{center}
\includegraphics[width=0.99\textwidth]{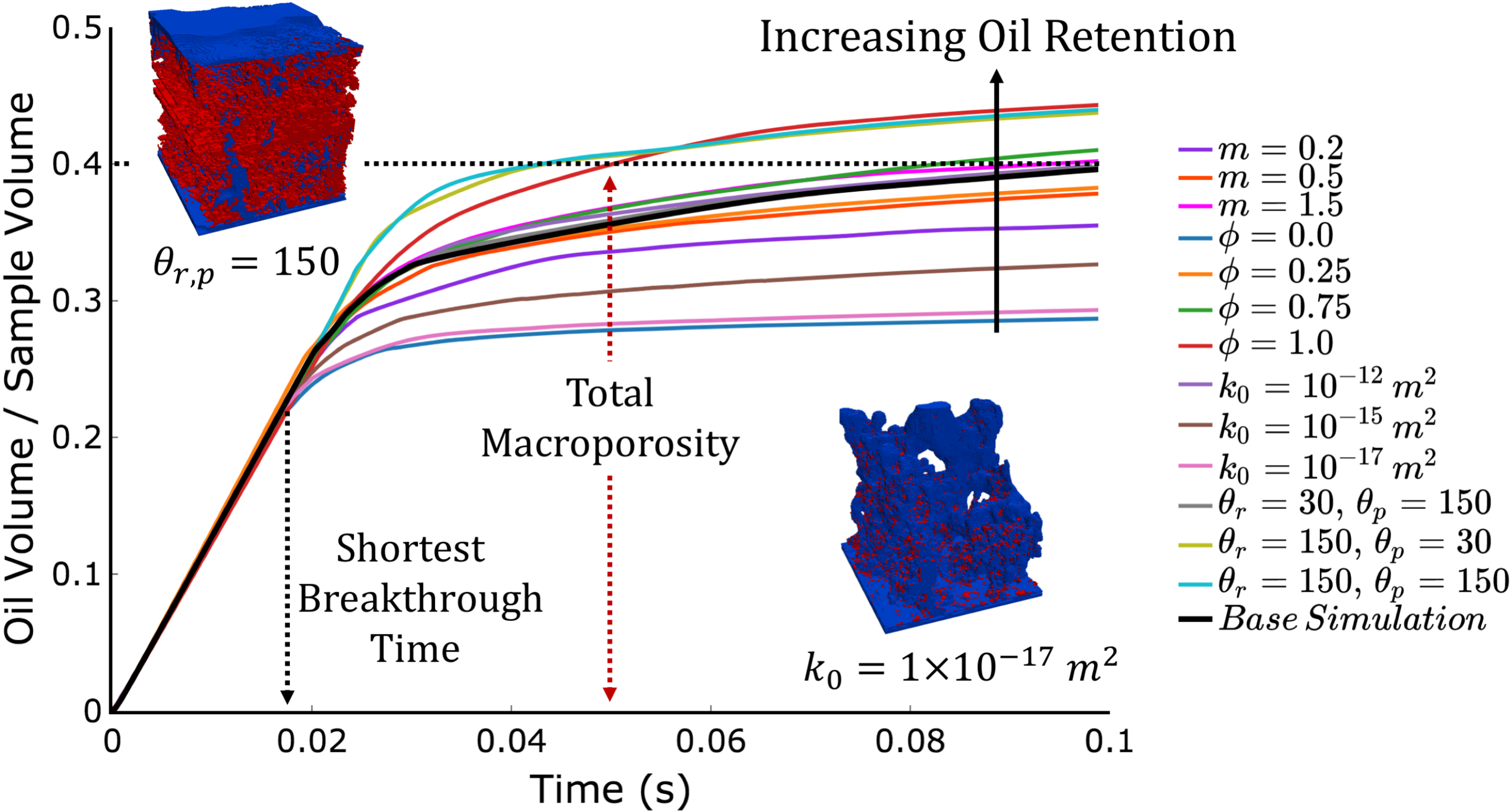}
\caption{\label{fig:graph_saturations} The evolution of oil saturation vs time for all studied cases. Each label shows the only varied parameter with respect to the base simulation. Note that oil breakthrough occurs when about $1/2$ of the macropore space still contains water, suggesting that much of the later (and slower) increase in oil content corresponds to water drainage from open pore space. The insets at the top left and bottom right show the final configurations for the cases with the highest and lowest final oil saturations, respectively. In said insets, the blue phase represents oil within open pore space and red represents the SRP that has been invaded by oil.}
\end{center}
\end{figure*}



Figure \ref{fig:graph_saturations} shows that fluid injection into the sample follows two characteristic behaviours: 1) An initial linear increase in saturation, where the slope is primarily dictated by the injection rate. 2) A non-linear plateauing slope dictated by the slow drainage of the defending fluid through the SRP and flow of the injected fluid into the SRP, which are influenced by the SRP's flow properties. The transition point between these two primary flow mechanisms is dictated by the ``breakthrough time", the point at which the injected fluid first reaches the sample's outlet boundary. The next two sections will leverage the information within the oil-flooding saturation curves in Figure \ref{fig:graph_saturations} to study the effects of the SRP on the dynamic and static properties of these experiments. 

\subsection{Impact on Breakthrough Time}

We now present a general ranking of the breakthrough times for oil flooding as a function of SRP properties obtained from the results in Figure \ref{fig:graph_saturations} (and Figure SI2). The samples are well distributed around the standard base case and obey the following trends: The slowest breakthrough times correspond to cases with oil-wetting surface contact angles, where the oil explores more of the porous medium before reaching the outlet, in agreement with experimental observations of multiphase flow in bead-packs and micromodels with no SRP \citep{Zhao2016,Hu2017}. These are followed by the sample case with no SRP, where the reasoning is the same as above. Sample cases with a less water-wet SRP (decreasing $m$) or with lower SRP permeability or porosity further decrease the breakthrough times by limiting the ability of water to drain through the SRP, such that the oil explores less of the sample before reaching the outlet. Overall, our results show that oil breakthrough times are sensitive to SRP parameters ($\pm \ 30\%$) even though drainage occurs predominantly in the larger pores, a result that has potentially important implications in enhanced oil recovery and geologic CO$_2$ sequestration.

\begin{figure}[ht!]
\begin{center}
\includegraphics[width=0.483\textwidth]{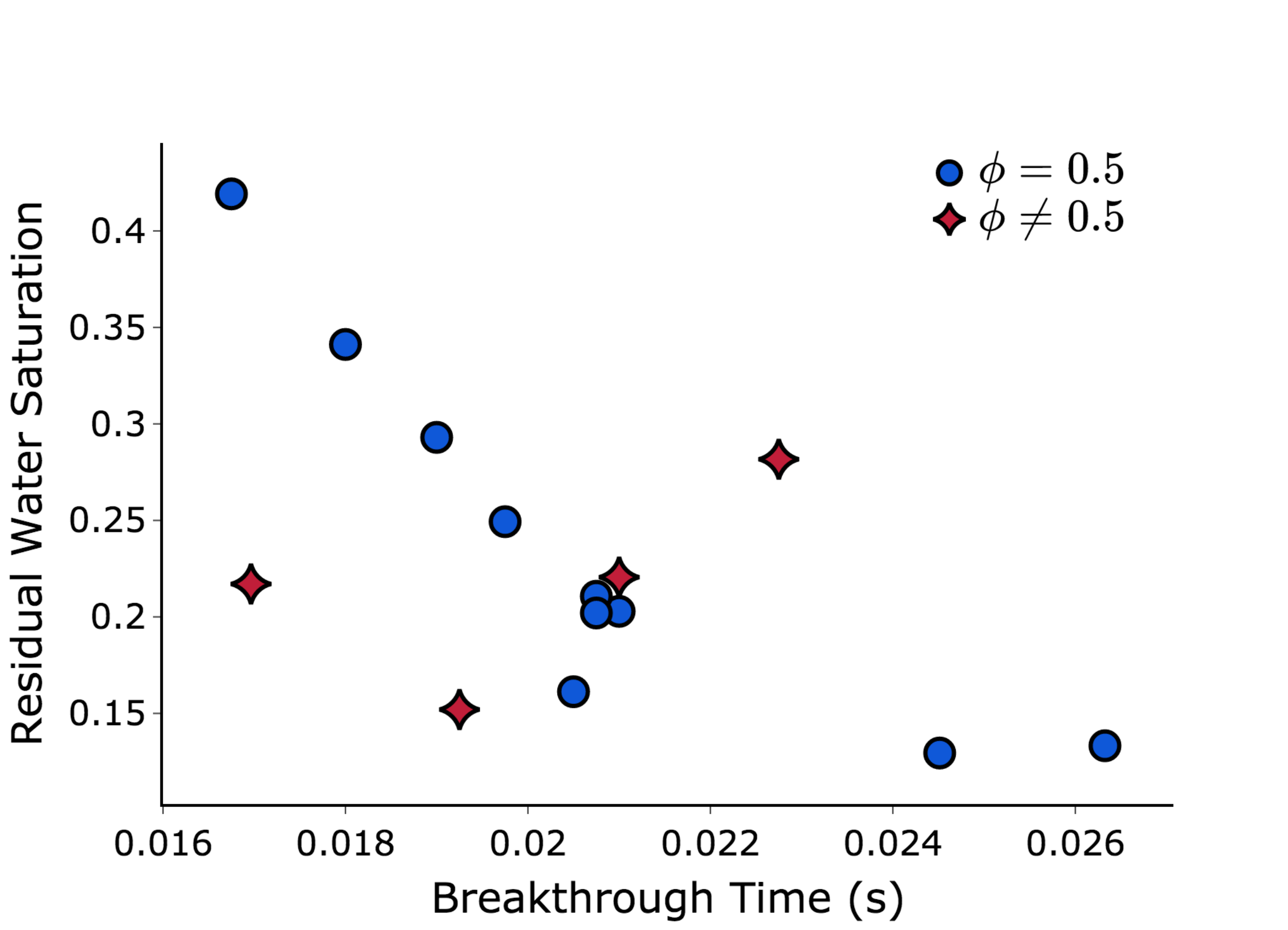}
\caption{\label{fig:sat_vs_time} Residual water saturations vs breakthrough times for all studied cases. Note that the longer it takes for oil to break through the sample, the lower the final water saturation. This trends holds for all cases with an SRP porosity of $\phi=0.5$ (blue circles). Changes in SRP porosity have the additional effect of changing the water storage capacity of the sample (red stars). 
}
\end{center}
\end{figure}

\subsection{Impact on Residual Saturations} 

Finally, we observe in Figure \ref{fig:sat_vs_time} (and Fig. SI3) that residual water saturations are highly correlated with oil breakthrough times: samples with faster breakthrough times generally have higher residual water saturations at steady state. The reasoning behind this behaviour is very similar to the one developed to explain the difference in oil breakthrough times: if more residual defending fluid is present, the invading fluid explores less of the available space and hence travels through the sample more rapidly. Overall, this analysis indicates that SRP has a considerable impact on a sample's residual saturations ($\pm 400\%$), strongly implying that it should not be neglected during the design of subsurface fluid extraction and sequestration processes.

\section{Conclusions}\label{sec:conclusion}

In this paper we studied the effects of XCT Sub-Resolution Porosity (SRP) on a rock's absolute permeability, relative permeability, residual saturations, and fluid breakthrough times. Our results quantify how these four properties react to changes in the porosity, permeability, and wettability of the SRP. One notable finding is that SRP can function as a persistent connector between otherwise-isolated fluid clusters during multiphase flow, even at low saturations. These results were obtained from numerical simulations performed with our newly-developed Multiphase Micro-Continuum framework. To the best of our knowledge, this is the first two-phase flow model and study to take into account SRP without having to rely on a quasi-static assumption or simplified pore-network models. 

As such, this investigation establishes a framework for performing two-phase flow simulations in digital rock systems that have two characteristic length-scales. Potential improvements to our methodology include the simulation of larger and more diverse rock samples, a very attainable task due to the current continuous and massive growth of high-performance computing. 
Finally, our results suggest potentially fruitful opportunities for future work aimed at quantifying the effects of SRP on upscaled capillary pressure curves, and broadening the investigated parameter space to different types of rocks involving different geometries, different amounts of SRP, and different SRP-induced connectivity. These avenues will more extensively test the conclusions presented in this study and lead the way towards greater understanding of multiscale rock physics and the development of more accurate and predictive upscaled permeability models. 


\paragraph*{Acknowledgements}

This work was supported by the National Science Foundation, Division of Earth Sciences, Early Career program through Award EAR-1752982. F.J.C. was additionally supported by a Mary and Randall Hack ‘69 Research Award from the High Meadows Environmental Institute at Princeton University. C.S was sponsored by the French Agency for Research (Agence Nationale de la Recherche, ANR) through the labex Voltaire ANR-10-LABX-100-01 and the grant FraMatI ANR-19-CE05-0002. We do not report any conflicts of interest. The code for the computational model used in this manuscript is archived at \url{https://doi.org/10.5281/zenodo.4013969} \citep{hybridPorousInterFoam_code} and can also be found at \url{https://github.com/Franjcf}. The Estaillades Carbonate rock sample was obtained from \cite{sample_dataset} through the \href{https://www.digitalrocksportal.org/}{Digital Rock Portal}.

\section{Nomenclature}

\nomenclature{$\rho_i$}{Density of phase $i$ ($\unit{kg/m^3}$) }%
\nomenclature{$\rho$}{Single-field fluid density ($\unit{kg/m^3}$) }%
\nomenclature{$\boldsymbol{U}$}{Single-field fluid velocity ($\unit{m/s}$)}%
\nomenclature{$\boldsymbol{U}_r$}{Relative fluid velocity ($\unit{m/s}$)}%
\nomenclature{$m$}{Van-Genuchten wettability parameter}%
\nomenclature{$p$}{Single-field fluid pressure ($\unit{Pa}$)}%
\nomenclature{$p_c$}{Average capillary pressure ($\unit{Pa}$) }%
\nomenclature{$\boldsymbol{S}$}{Single-field fluid viscous stress tensor ($\unit{Pa}$)}%
\nomenclature{$Q$}{Volumetric fluid flow rate ($\unit{m^3/s}$)}%
\nomenclature{$\Delta P$}{Macroscopic pressure difference ($\unit{Pa}$)}%
\nomenclature{$\gamma$}{Fluid-fluid interfacial tension ($\unit{Pa.m}$)}%
\nomenclature{$\phi$}{Porosity field}%
\nomenclature{$\alpha_w$}{Saturation of the wetting phase}%
\nomenclature{$\alpha_n$}{Saturation of the non-wetting phase}%
\nomenclature{$\mu_i$}{Viscosity of phase $i$ ($\unit{Pa.s}$) }%
\nomenclature{$k_0$}{SRP absolute permeability ($\unit{m^2}$) }%
\nomenclature{$K_0$}{Sample absolute permeability ($\unit{m^2}$) }%
\nomenclature{$k_{r,i}$}{SRP relative permeability for fluid $i$ }%
\nomenclature{$K_{r,i}$}{Sample relative permeability for fluid $i$ }%
\nomenclature{$\boldsymbol{F}_{c}$}{Average capillary forces ($\unit{Pa/m}) $}%
\nomenclature{$C_{\alpha}$}{Parameter for the compression velocity model}%
\nomenclature{$M_i$}{Mobility of phase $i$ ($\unit{m^3/kg.s}$)}%
\nomenclature{$M$}{Total mobility ($\unit{m^3/kg.s}$)}%
\nomenclature{$\theta_r$}{Rock surface contact angle}%
\nomenclature{$\theta_p$}{SRP surface contact angle}%
\nomenclature{$\boldsymbol{n}_{wall}$}{Normal vector to the porous surface}%
\nomenclature{$\boldsymbol{t}_{wall}$}{Tangent vector to the porous surface}%
\nomenclature{$p_{c,0}$}{Entry capillary pressure ($\unit{Pa}$)}%
\nomenclature{$V_f$}{Total volume of fluid in the sample ($\unit{m^3}$) }%
\nomenclature{$V_{f,i}$}{Total volume of fluid $i$ in the sample ($\unit{m^3}$) }%

\printnomenclature\label{nom}

\bibliographystyle{chicago}
\bibliography{bibliography,ref}

\end{document}